\documentclass{Interspeech}

\usepackage{multicol,multirow}
\usepackage{arydshln}

\interspeechcameraready

\title{Improving Large-Scale Weakly Supervised ASR by Filtering and Selection}

\author[affiliation={1,2}]{Kohei}{Matsuura}
\author[affiliation={1}]{Masato}{Mimura}

\affiliation{Human Informatics Laboratories}{NTT Corporation}{Japan}
\affiliation{Graduate School of Informatics}{Kyoto University}{Japan}
\email{kohei.matsuura@ntt.com}

\keywords{end-to-end automatic speech recognition, 
    large-scale weakly supervised pretraining, 
    data filtering and selection
}

\usepackage{comment}

\begin{document}

\maketitle
\begin{abstract}
Leveraging large-scale weakly supervised datasets is crucial to train robust end-to-end automatic speech recognition (ASR) models.
However, such datasets often contain noisy labels and lack domain specificity, limiting their effectiveness.
To address these issues and make better use of weakly supervised datasets, we propose a novel training approach incorporating data filtering and selection.
Our approach consists of three steps: pretraining on the entire dataset, continued pretraining on a filtered subset based on character error rate (CER), and fine-tuning on a small number of acoustically similar samples to the target domain, selected from the filtered subset.
In experiments with a 90,000-hour weakly supervised Japanese dataset, the proposed filtering and selection methods synergistically reduced CER by up to 6.4\% and 4.0\%, respectively, even though these steps reused training samples already used in the first pretraining step.
\end{abstract}

\section{Introduction}
End-to-end automatic speech recognition (ASR), which converts spoken utterances into transcriptions using a single deep learning model, has seen remarkable progress over the past five years~\cite{prabhavalkar2024}.
In particular, large-scale self-supervised~\cite{baevski2020,abdelrahman2022} and weakly supervised~\cite{zhou2017,radford2023} pretraining has played a key role.
Self-supervised learning (SSL) leverages unlabeled speech to learn general speech representations and is applicable beyond ASR.
Meanwhile, weakly supervised learning, which we focus on in this study, relaxes the need for manually verified and precise labels when collecting task-specific training samples typically from internal speech archives and the Internet.
Since it does not require expensive high-quality transcriptions, we can obtain  very large-scale ASR datasets that encompass diverse speakers, speaking styles, and background noises.

Although weakly supervised learning is effective for ASR, it has two potential issues.
One is the existence of low-quality labels, such as misspellings and unspoken supplementary text.
These noisy labels may negatively affect ASR accuracy.
In semi-supervised learning, where an ASR model generates labels for unlabeled speech to retrain itself, it is common practice to filter out noisy labels considering confidence scores~\cite{khan2020}. 
However, the effectiveness of such filtering strategies remains largely under-explored on weakly supervised datasets.
The other issue stems from the diversity of these datasets. 
Trained with vast and diverse samples, ASR models may remain suboptimal for specific target domains.
To alleviate this problem, fine-tuning the pretrained ASR models using samples similar to those in the target domains (i.e., data selection) is promising.
Nevertheless, most studies have investigated data selection in the context of core-set selection~\cite{xiong2020,lu2022}, where the primary motivation is not domain adaptation but reducing the training data.
Few studies tackle on data selection to further improve robust pretrained models using weakly supervised datasets.

\begin{figure}[t!]
    \hspace{10pt}
    \centering
    \includegraphics[width=0.48\textwidth]{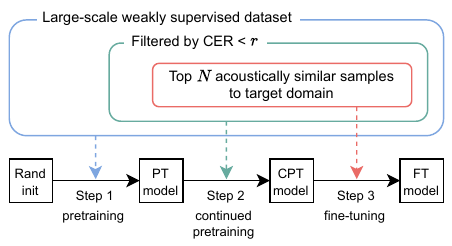}
    \caption{Overview of our proposed approach, which appropriately narrows a single weakly supervised dataset to gradually improve an ASR model. PT, CPT, and FT denote pretrained, continually pretrained, and fine-tuned models, respectively.}\vspace{-5pt}\vspace{-5pt}
    \label{fig:overview}
\end{figure}

To address these issues and fully exploit weakly supervised datasets, we propose a novel training approach incorporating data filtering and data selection.
Our approach, as depicted in Figure~\ref{fig:overview}, involves the following three steps:
(1) First, we pretrain an encoder-only ASR model based on connectionist temporal classification (CTC) \cite{graves2006} with the entire large-scale weakly supervised dataset.
Then, we use this pretrained model to transcribe all speech in the dataset and compute a character error rate (CER) between the predicted and the original noisy labels to each sample.
(2) We filter out samples with CER higher than a predefined threshold $r$ and continue pretraining the model using the remaining data.
(3) If a target-domain training set is not available, we fine-tune the continually pretrained model obtained in Step 2 by selecting samples acoustically similar to the target-domain speech.
Since complex similarity measures are not practical for large-scale datasets, we propose a simple method using cosine similarity on speech SSL embeddings.

To evaluate our approach, we conducted experiments on a 90,000-hour Japanese weakly supervised ASR dataset, collected following~\cite{li2023}, and three public evaluation datasets.
While most previous studies utilize such datasets only for Step 1 (i.e., simple pretraining), our experiments demonstrated further significant improvements through both Step 2 and 3.
We also found that the filtering threshold $r$ should be neither too low nor too high, revealing a trade-off between label quality and data diversity.
Moreover, this improvement and trade-off trend persisted after fine-tuning when target-domain training sets were available, suggesting that our approach is widely applicable.

\section{Related Work}
\subsection{Large-scale ASR pretraining and data filtering}
\label{sec:related_filtering}
In recent years, ASR models trained on very large datasets, such as Whisper~\cite{radford2023}, OWSM~\cite{peng2024a}, and Canary~\cite{puvvada2024}, have emerged.
Notably, Whisper achieves high robustness across diverse speech inputs by utilizing large-scale weakly supervised datasets collected from the Internet.
However, weakly supervised datasets lack careful annotations and contain noisy labels with misspellings, word omissions or insertions due to misaligned timestamps, and unspoken supplementary text.
Despite noisy labels potentially degrading recognition accuracy, the Whisper models are trained on most of the collected data with only minor heuristic filtering.

Several studies have explored extensive data filtering.
\cite{galves2021} applied CER-based filtering, removing samples with a CER above 50\% using a transducer ASR model; however, they did not experimentally validate its effectiveness.
\cite{li2023} evaluated CTC score-based filtering, but their analysis was limited to a 160-hour dataset. 
Hence, the impact of filtering on large-scale weakly supervised datasets remains largely unexplored.

\subsection{Data selection from large-scale datasets}
\label{sec:related_adaptation}
In ASR research, data selection from large-scale datasets has been widely studied in the context of \textit{core-set selection}.
Core-set selection aims to reduce training time by selecting a small subset, while maintaining recognition accuracy comparable to that obtained using the full dataset.
Siohan et al. proposed a greedy method that minimizes the Kullback-Leibler divergence between the i-vector distributions of the selected and target dataset~\cite{siohan2013}.
Subsequently, \cite{asami2015} applied submodular optimization~\cite{shinohara2014} for efficient selection and replaced i-vectors with Gaussian Mixture Models (GMMs) of acoustic features.
More recently, \cite{lu2022} leveraged n-gram language models trained on discrete representations from w2v-BERT~\cite{chung2021} or VQ-VAE~\cite{oord2017}.

A few studies have explored data selection to improve recognition accuracy in the target domain.
\cite{mimura2014} selected speech from speakers similar to the target using Euclidean distance and cosine similarity of GMM super-vectors for speaker adaptation.
\cite{kothawade2023} efficiently selected speech similar to the target accent using submodular mutual information functions.
The most relevant work is by Lagos et al., who fine-tuned a pretrained ASR model using samples selected via SSL embeddings for domain adaptation~\cite{lagos2024}.
However, their methods differ from ours in that they used separate datasets for pretraining and data selection, which may limit its applicability when an additional dataset is unavailable.
Moreover, their dataset for data selection is only one-tenth the size of ours.

\section{Methodology}
\label{sec:proposed}

\subsection{Pretraining (Step 1)}
\label{sec:proposed_pretraining}
This section describes the proposed training approach on large-scale weakly supervised dataset.
First, we pretrain an CTC-based encoder-only ASR model using the entire dataset.
While this dataset contains noisy labels, the majority of the dataset contains labels that reflect the speech content.
Thus, the impact of noise is expected to be mitigated during a large-scale training, allowing the model to coarsely learn the ASR task.

The primary reason for adopting a simple CTC-based ASR model is to perform large-scale filtering efficiently in the next section.
Furthermore, it also offers several other advantages.
\cite{peng2024b} argues that CTC-based ASR models are robust to hallucination, and therefore we expect it is suitable for weakly supervised learning.
In addition, we can use it to initialize the encoder components of more advanced ASR models, such as encoder-decoder \cite{vaswani2017} or transducer \cite{graves2012,gulati2020} models.

\begin{table}[t]
    \centering
    \caption{Original (``orig''), predicted (``pred''), and Oracle labels with CERs (\%), translated from Japanese.}
    \begin{tabular}{c|c|cl}
    \hline
         ID & CER (\%)&  \multicolumn{2}{c}{labels}\\
    \hline
         \multirow{3}{*}{1}& \multirow{2}{*}{25.0} & orig & \textbf{but whe}re the knee bends\\
            &   & pred & are the knee bends \\\cdashline{2-4}
            &   & Oracle & \textit{...re the knee bends} \\
    \hline
         \multirow{3}{*}{2}& \multirow{2}{*}{11.5} & orig & using lightly grilled sardine \\
            &   & pred & using lightly grilled \textbf{starting} \\\cdashline{2-4}
            &   & Oracle & \textit{using lightly grilled sardine} \\
    \hline
    \end{tabular}\vspace{-5pt}\vspace{-5pt}
    \label{table:filtering_samples}
\end{table}

\subsection{Filtering and continued pretraining (Step 2)}
\label{sec:proposed_filtering}
Next, we transcribe all speech in the dataset using the CTC-based ASR model obtained in Step 1 and calculate the CER between the predicted and original weak labels.
We regard samples with low CERs as having high-quality labels.

The ASR model used to compute CERs needs to be robust across various domains; otherwise, the model may assign a high CER to samples from unfamiliar domains even though their labels are correct.
In this study, since CER is calculated on the same dataset used for training, the model is expected to recognize all samples robustly.
In addition, given the very large scale of the training data, the model is unlikely to accurately memorize the noisy labels.
Instead, it is expected to transcribe the input speech faithfully based on the spoken content.
Thus, we can detect noisy labels using CERs to some extent.
Table~\ref{table:filtering_samples} shows actual examples of the ASR model's predictions, the original weak labels, and the Oracle spoken content.
In ID~1, since the ASR model did not transcribe the unspoken portions, the CER was high, correctly indicating that the original label was noisy.
In contrast, like ID~2, ASR predictions may contain errors, and inaccurately high CERs are sometimes observed.
Note that this method does not perfectly distinguish between low-quality labels and recognition errors.
However, our empirical findings suggest that our method is particularly useful for detecting deletions and insertions by misalignment of time-stamps in the original labels.

Then, we filter out samples with CERs higher than a threshold $r$ and continue pretraining using the remaining samples.
A low filtering threshold $r$ will retain high-quality labeled samples.
However, there is a risk of filtering out samples that are difficult to recognize correctly, despite they have accurate labels.
If such hard-to-recognize samples are removed and the training data becomes biased toward easier samples, the speech recognition model may lose its robustness.
Hence, choosing the appropriate threshold $r$ is important. 
In this study, we vary the threshold $r$ to analyze the trend in final recognition accuracy.
Furthermore, if a target-domain training set is available, we examine whether the trend persists after fine-tuning.
Since pretrained models are typically fine-tuned before use, this analysis is essential for obtaining practical insights.

\subsection{Data selection and fine-tuning (Step 3)}
\label{sec:proposed_adaptation}
Finally, if a target-domain training set is not available, we select samples acoustically similar to the target-domain speech and fine-tune the continually pretrained model obtained in Step~2.

We select similar samples as follows.
First, each speech~$\mathbf{x}$ in the target-domain dataset $\mathcal{D}_{tgt}$ is embedded using a SSL model $\mathrm{Emb(\cdot)}$, time-averaged to obtain a single vector representation, and then averaged across all samples to compute the mean vector $\mathbf{a}$, which represents the center of the target domain.
\begin{eqnarray}
\mathbf{a} = \frac{1}{|\mathcal{D}_{tgt}|}\sum_{x \in \mathcal{D}_{tgt}} \frac{1}{L}\sum_{1 \leq l \leq L}\mathrm{Emb}(\mathbf{x})_l,
\end{eqnarray}
where $L$ denotes the length of $\mathrm{Emb}(\mathbf{x})$, and $\mathrm{Emb}(\mathbf{x})_l$ represents the $l$-th vector of the embedded $\mathbf{x}$.
Next, the same SSL model embeds each speech $\mathbf{x'}$ in the weakly supervised dataset $\mathcal{D}$, and we average the embeddings along the time axis to obtain a single vector $\mathbf{e}$.
\begin{eqnarray}
    \mathbf{e} &=& \frac{1}{L'}\sum_{1 \leq l' \leq L'}\mathrm{Emb}(\mathbf{x'})_{l'},
\end{eqnarray}
where $L'$ denotes the length of $\mathrm{Emb}(\mathbf{x'})$.
Then, the cosine similarity \( s \) is computed between each \(\mathbf{e}\) and the mean vector \(\mathbf{a}\).
\begin{eqnarray}
    s &=& \mathrm{cos\_sim}(\mathbf{a}, \mathbf{e}) =  \frac{\mathbf{a} \cdot \mathbf{e}}{\lVert\mathbf{a}\rVert\lVert\mathbf{e}\rVert}.
\end{eqnarray}
Finally, the \( N \) samples with the highest similarity \( s \) are selected for fine-tuning.
This method leverages expressive speech representations from SSL models, 
while maintaining linear complexity and easily scaling to even larger datasets.

\begin{figure}[t!]
    \hspace{-11pt}
    \centering
    \includegraphics[width=0.48\textwidth]{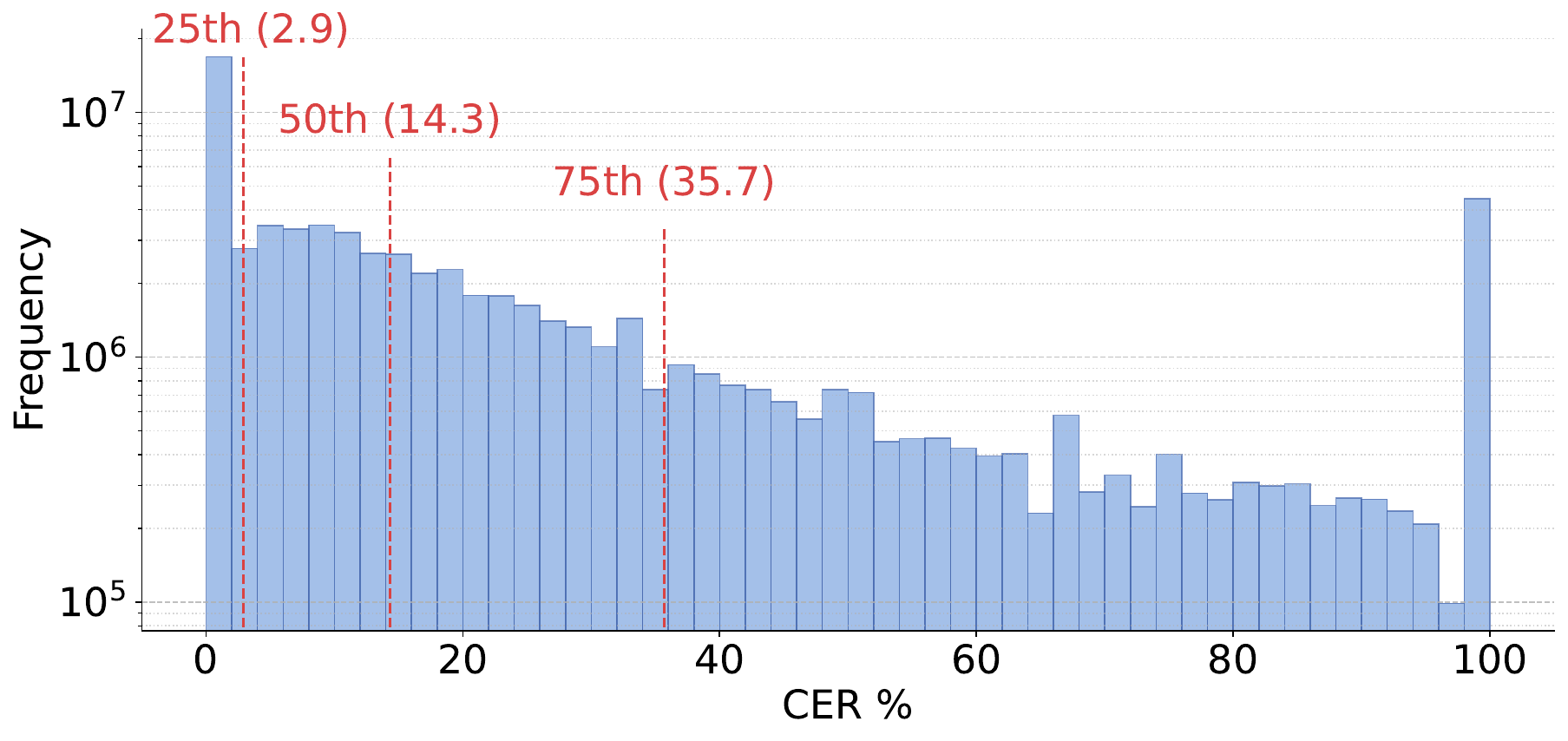}
    \caption{CER distribution of the collected dataset calculated between the original noisy labels and predictions of the pretrained model, with the 25, 50, and 75th percentile values.}\vspace{-5pt}
    \label{fig:cer_distribution}
\end{figure}

\section{Experimental Evaluation}
\label{sec:evaluation}
\subsection{Datasets}
\label{sec:datasets}
To validate the effectiveness of the proposed approach, we collected a large amount of speech and Japanese transcription pairs from the Internet following~\cite{li2023}.
We used an external language identification model\footnote{{HuggingFace: speechbrain/lang-id-voxlingua107-ecapa}} trained on the VoxLingua107 dataset~\cite{valk2021} to filter out samples with non-Japanese speech.
We also removed samples shorter than 1 second or longer than 30 seconds.
Additionally, we performed basic text preprocessing, including removing parentheses, normalizing characters, and eliminating unpronounced symbols.
As a result, we obtained a large-scale Japanese weakly supervised dataset containing approximately 90,000 hours of speech (72M samples).
Figure~\ref{fig:cer_distribution} illustrates the distribution of CERs, capped at 100\%, calculated as described in Section~\ref{sec:proposed_filtering}.
The 25\% of the dataset consisted of samples with CERs of 2.9\% or lower, suggesting that they contained almost accurate labels.
On the other hand, over 3M samples had CERs of 100\% or higher, indicating that the original labels did not reflect the corresponding spoken content at all.

For evaluation, we adopted three public Japanese ASR datasets: the Corpus of Spontaneous Japanese (CSJ)~\cite{maekawa2003}, CommonVoice (CV)~\cite{ardila2020}, and Noisy-KU\footnote{{GitHub: Kyoto-University-Speech-and-Audio/noisy-csj}}.
\textbf{CSJ} is a high-quality dataset that includes academic lectures and everyday interviews.
The training and evaluation sets contain 660 hours and 5.1 hours of speech, respectively.  
There are three evaluation subsets, and we report the average CER across them.
\textbf{CV} is a multilingual ASR dataset, where many anonymous volunteers recorded sentences using various microphones and environments.
We used the Japanese subset of version 17.
The training and evaluation sets contain 13.5 and 8.9 hours of speech, respectively.
\textbf{Noisy-KU} includes controlled spontaneous monologues recorded in cafes, museums, and roadside with strong background noise.
It consists of a 2.1-hour evaluation set without a training set.
We used distant speech recorded with an array microphone.

\begin{figure*}[t]
    \centering
    \includegraphics[width=1.0075\linewidth]{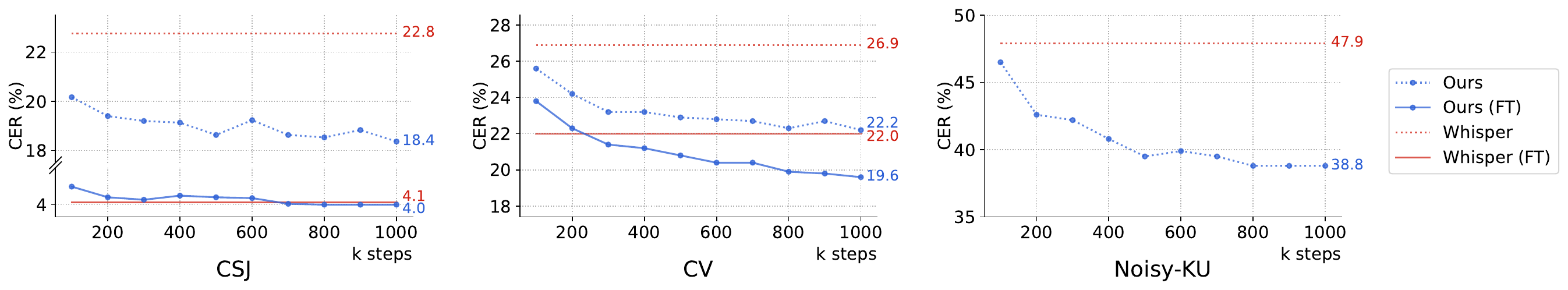}
    \captionsetup{width=0.9\textwidth}
    \caption{CERs ($\downarrow$) transition during pretraining, with fine-tuning (``FT'') every 100k steps for datasets with training sets.}\vspace{-5pt}
    \label{fig:result_pret}
\end{figure*}

\begin{figure*}[t]
    \centering
    \includegraphics[width=1.0075\linewidth]{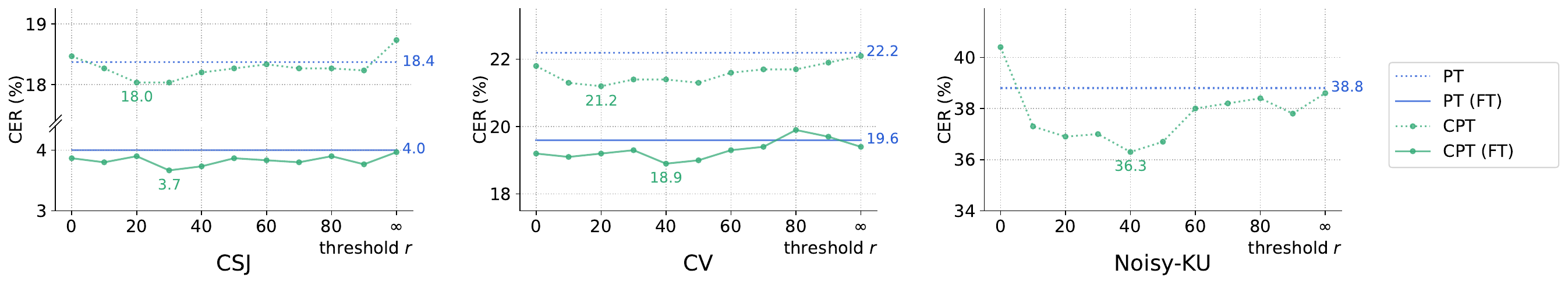}
    \captionsetup{width=0.9\textwidth}
    \caption{CERs ($\downarrow$) after 1M-step pretraining (``PT'') and additional 100k-step continued pretraining with various filtering thresholds \( r \) (``CPT''), along with the best CER values. Fine-tuning (``FT'') was applied for datasets with training sets.}
    \label{fig:result_cpret}\vspace{-5pt}
\end{figure*}

\subsection{ASR Models}
As described in Section~\ref{sec:proposed_pretraining}, we trained a CTC-based encoder-only model.
It consisted of 17-layer Conformer blocks \cite{gulati2020} with an attention dimension of 512, 8 attention heads, and a kernel size of 15, replacing all batch normalization with layer normalization following~\cite{matsuura2022}.
The vocabulary consisted of the most frequent 5,000 characters in the Japanese subset of the CC-100 dataset~\cite{wenzek2020}.
The ASR model had 117.3M parameters.

We also evaluated the Whisper Small ASR model~\cite{radford2023} for reference.
The Whisper model was trained on a 670,000-hour multilingual weakly supervised dataset, which included 7,054 hours of Japanese speech-transcription pairs.
This model consisted of a 12-layer Transformer encoder and a 12-layer Transformer decoder~\cite{vaswani2017} with 240.6M parameters, which was twice the size of our model.

\subsection{Detailed settings}
In Step 1, we pretrained the model for 1M steps.
The learning rate was linearly increased from 0.0 to \(1.0 \times 10^{-4}\) over the first 10,000 steps and then kept constant with Adam optimizer~\cite{kingma2015}. 
Each mini-batch contained 1,000 seconds of speech (averagely 222.1 samples).
Since the whole training dataset consisted of about 72M samples, the model was trained for about 3 epochs.
Pretraining was performed using 4 NVIDIA RTX A6000 GPUs and required 507 hours.  
We saved checkpoints every 1,000 steps.
To report CERs, we averaged the last 10 checkpoints and used greedy decoding at the frame level.

For continued pretraining in Step 2, we filtered the entire dataset with CER thresholds $ r = \{0, 10, 20, 30, 40,$ $ 50, 60, 70, 80, 90, \infty\} $.
Here, \( r = \infty \) denotes that no filtering was applied.  
Then, we continued training the model obtained in Step 1 using each filtered dataset for an additional 100k steps.  
During Step 2, the learning rate was kept constant at \(1.0 \times 10^{-4}\), and all other training settings remained the same as in Step 1.

For evaluation on the CSJ and CV datasets, we then fine-tuned the continually pretrained model using their training sets.  
During fine-tuning, we set the learning rate to \(5.0 \times 10^{-5}\) and used the WarmupLR scheduler with 2,000 warmup steps.  
We evaluated the validation set every 1,000 steps, and fine-tuning was stopped if the CER did not improve for five consecutive evaluations.  
The final model was obtained by averaging the last 5 checkpoints.  
In most cases, early stopping occurred within 20,000 steps for CSJ and 10,000 steps for CV.

For the Noisy-KU dataset, we applied the data selection methods described in Section~\ref{sec:proposed_adaptation} (i.e., Step 3).  
Specifically, we selected the top 500k samples with the highest similarity scores \( s \) from datasets filtered at thresholds \( r = \{0, 10, 20, 30\} \), using a Japanese HuBERT base model~\cite{hsu2021,sawada2024}.
To validate the effectiveness of this method, we created four additional training sets by randomly selecting 500k samples from each filtered dataset.  
Using these datasets, we fine-tuned the continually pretrained model with \( r = 30 \) from Step 2 under the same settings as CSJ and CV\footnote{Since we should not determine $r$ using the Noisy-KU evaluation set, we selected the best $r$ based on the CSJ and CV validation sets.}.  
Since Noisy-KU lacks a validation set for early stopping, the number of training steps was fixed at 10,000.

For the Whisper model, we halved the batch size, set the learning rate to \(1.0 \times 10^{-5}\), and applied SpecAugment~\cite{park2019}, based on the preliminary experiments on the validation sets of CSJ and CV. 
For inference, we used beam search with a beam width of 8.
The other fine-tuning settings remained the same.

\begin{table}[t]
    \centering
    \renewcommand{\arraystretch}{1.2}
    \setlength{\tabcolsep}{4.5pt} \vspace{2pt}
    \caption{CERs ($\downarrow$) on Noisy-KU after fine-tuning (``FT'') using 500k samples selected by acoustic similarity (``sim'') or randomly (``rand'') from filtered datasets with threshold $r$.}
    \begin{tabular}{c|cccc:c}
    \hline
        $r$ & 0 & 10 & 20 & 30 & w/o FT \\
    \hline
        sim & 37.9\scriptsize{$\pm$0.3} & 36.5\scriptsize{$\pm$0.4} &       35.7\scriptsize{$\pm$0.3} & \textbf{35.3}\scriptsize{$\pm$0.1} & \multirow{2}{*}{37.0} \\
        rand  & 38.4\scriptsize{$\pm$0.2} & 37.3\scriptsize{$\pm$0.3} &       36.8\scriptsize{$\pm$0.2} & 36.7\scriptsize{$\pm$0.3} & \\
    \hline
    \end{tabular}\vspace{-5pt}\vspace{-5pt}
    \label{tab:result_adaptation}
\end{table}

\subsection{Results}
Figure~\ref{fig:result_pret} shows CERs for each dataset at every 100k steps during the pretraining in Step 1.  
Since the Whisper model is not specialized for Japanese, the CERs remained relatively high without fine-tuning on the CSJ or CV training sets.  
However, even after fine-tuning, our simple CTC-based model (``Ours'') outperformed it, indicating that our experiment was conducted at a state-of-the-art level.

The CERs consistently decreased as training progressed.  
Even when trained with noisy labels, the model gradually learned the ASR task.
These results also suggest that a 100M-parameter model still retained unused capacity after one epoch on the 90,000-hour dataset.

Figure~\ref{fig:result_cpret} presents the results of filtering and continued pretraining with various CER thresholds \( r \) in Step 2.  
Notably, we observed consistent improvement across all datasets, despite reusing filtered but already seen training samples.
The optimal threshold was found to be within the range of \( 20 \leq r \leq 40 \) across the dataset, and we gained up to 6.4\% relative CER reduction.
Interestingly, we gained limited improvement with presumably clean labels at $r \leq 10$.
This suggests that balancing label quality and data diversity is important to effectively improve recognition accuracy with weakly supervised datasets.
Furthermore, since this trend persisted after fine-tuning on the CSJ and CV training sets, our filtering method and findings are useful in practical scenarios.

Table~\ref{tab:result_adaptation} presents the results of domain adaptation in Step 3 on the Noisy-KU dataset.
Compared to fine-tuning with randomly selected data, selecting acoustically similar samples resulted in significant improvements in all settings.
During fine-tuning, maintaining diversity was also crucial, and a threshold of \( r = 30 \) yielded the lowest CER of 35.3\%, achieving a relative CER reduction of 4.0\% compared to the model without Step 3. 
To the best of our knowledge, this is the first study to further improve a robust pretrained ASR model by selecting an appropriate subset from the same dataset used for pretraining.

\section{Conclusion}
In this study, we proposed an approach to effectively exploit large-scale weakly supervised datasets for Japanese ASR, focusing on data filtering and domain adaptation via data selection.
Our experiments demonstrated that applying CER-based filtering enhanced label quality and improved ASR accuracy, while maintaining data diversity was also crucial.
Additionally, we showed that fine-tuning with acoustically similar samples was effective for domain adaptation, despite using a subset of the same dataset from pretraining and continued pretraining.
Finally, our approach reduced CER by 9.0\% (from 38.8\% to 35.3\%) on the Noisy-KU dataset solely by appropriately narrowing and reusing a single training dataset.


\bibliographystyle{IEEEtran}
\bibliography{refs}

\end{document}